\documentclass[aps,preprint,onecolumn,showpacs,preprintnumbers,groupedaddress,nofootinbib]{revtex4}
\usepackage{graphics}  
\usepackage{epsfig}
 \usepackage{verbatim}

\IfFileExists{srcltx.sty}{\usepackage[active]{srcltx}}

\newcommand{\be}{\begin{equation}}
\newcommand{\ee}{\end{equation}}
\newcommand{\bea}{\begin{eqnarray}}
\newcommand{\eea}{\end{eqnarray}}

\begin{document}

\title{A gamma-ray signature of energetic sources of cosmic-ray nuclei}

\author{Alexander Kusenko}
\affiliation{Department of Physics and Astronomy, University of California, Los 
Angeles, CA 90095-1547, USA}
\affiliation{IPMU, University of Tokyo, Kashiwa, Chiba 277-8568, Japan}
 \author{M.B. Voloshin}
\affiliation{William I. Fine Theoretical Physics Institute, University of Minnesota,
Minneapolis, MN 55455, USA}
\affiliation{
Institute of Theoretical and Experimental Physics, Moscow, 117218, Russia}



\begin{abstract}
Astrophysical sources of nuclei are expected to produce a broad spectrum of isotopes, many of which are unstable.  An unstable nucleus can 
beta-decay outside the source into a single-electron ion.  Heavy one-electron ions, thus formed, can be excited in their interactions 
with cosmic microwave background photons, in which case they relax to the ground state with the emission of a gamma ray. 
Repetitive cycles of excitation and gamma-ray emission can produce an observable feature in the gamma-ray spectrum with a maximum around 8~GeV (for iron). We find that the observed spectrum of Centaurus A is consistent with a substantial flux of nuclei accelerated to 0.1~EeV.  
A characteristic 5--10~GeV  (iron) shoulder in the gamma-ray spectra of various sources can help identify astrophysical accelerators of nuclei or set upper limits on nuclear acceleration.
\end{abstract}

\maketitle

Heavy nuclei are detected in high-energy cosmic rays~\cite{Bluemer:2009zf}, although the evidence is still controversial at the highest end of the spectrum~\cite{Abraham:2009dsa}. The sources 
are difficult to identify because the arrival directions are altered by the galactic magnetic fields.  It would be very desirable to identify 
astrophysical nuclear accelerators by some other means, and some signatures in gamma rays and neutrinos have been proposed~\cite{signatures}.  We describe a new 
signature that can be used to identify powerful sources of cosmic-ray nuclei: a characteristic ``shoulder'' in the spectrum of gamma rays.

Astrophysical sources, such as jets of active galactic nuclei (AGN), produce cosmic rays in an environment of high photon density, which can 
cause some accelerated nuclei to disintegrate.  However, a non-negligible fraction of nuclei can escape~\cite{survival}.  Gamma-ray data 
indicate that acceleration of cosmic rays does take place in AGNs: line-of-sight interactions of cosmic rays provide a natural 
explanation for the observed spectra of distant blazars~\cite{Essey:2009zg}, assuming the intergalactic magnetic fields are relatively small~\cite{Ando:2010rb}.  Models allow the photon density in the acceleration 
region to vary from a  low enough for the nuclei to escape to a high enough for photodisintegration~\cite{survival}.  
It is reasonable to assume that nuclei coming out of the source are subject to some degree of photodisintegration, which would render 
them generally unstable.  On average, it takes one beta decay for an unstable nucleus to become stable.  

Astrophysical sources can produce a number of different isotopes.  Unstable nuclei decay, mostly, via beta decays. We will use iron as an example. For a fully ionized large nucleus, the beta-decay electron is created in a bound state with probability that is of the order of one~\cite{Bahcall:1961zz}.  Hence, an order-one fraction of nuclei turn into one-electron ions at distance $c \tau_{\rm decay}\gamma$ from the source, where $\gamma=E/M$, $E$ is the nucleus energy, and $M$ is its mass.  Once a beta electron is bound in an ion, it is likely to remain bound even in the event of a subsequent beta decay, because the nuclear recoil and the relative change in the charge $Z$ of the nucleus are both small. 

The relevant energy corresponds to $\gamma \sim 10^6 - 10^7$, for which the cosmic microwave background (CMB) photons excite the electron from its ground state to one of the higher bound states.  The CMB photon energy in the frame of reference of the ion is $\approx 7$~keV for iron.  We will see that the ion is ionized, on average, about one year or more after it was created in a bound beta decay.  De-excitation of the excited state, which is practically instantaneous, produces a photon with energy $E_{\gamma,\rm i}\approx 7$~keV in the frame of the ion, which corresponds to energy $E_{\gamma,\rm lab}\sim \gamma E_{\gamma,\rm i}$ in the laboratory (CMB) frame. Repeated cycles of excitation and emission can create a gamma-ray signal that can help 
identify the sources of UHECR nuclei. We note that the probability of a beta electron to be captured on an excited $nS$ atomic level (with $n\ge2$) is approximately 0.2 of the capture probability for the ground state ($n=1$)~\cite{Bahcall:1961zz}. The de-excitation of these higher levels thus creates approximately 0.16 gamma photons per ion at the time of beta decay, before the re-excitation of the ion in collisions with the CMB.

The observable flux of gamma-rays emerging from the described process of excitation and de-excitation of the one-electron ions depends on the excitation rate $\Gamma_{\rm e}$ of the ion from the ground state to a discrete excited level due to interactions with CMB photons, and the rate of the photo-ionization due to interactions with CMB photons ($\Gamma_{\rm i}$) or extragalactic background light (EBL) photons ($\Gamma_{\rm i; \, EBL}$). The former process generates the gamma-ray flux, while the rate of ionization determines the lifetime of the ions in the photon bath. 
In what follows we refer to the laboratory (CMB) frame values of the rates.  The rates can be calculated using the dipole approximation for the radiative transitions, since at $Z \approx 26$ (iron) the atomic dynamics is still sufficiently nonrelativistic. In this approximation only the excitation of the ground state to $P$-wave states, discrete and in the continuum, is relevant. For the transition from the ground $1S$ to  a specific $nP$ state the rate in a photon bath can be readily found in the CMB frame in the form 
\bea
&&\Gamma_{{\rm e},n}=2 \,  \int n_{\rm CMB}(\omega) \, {d^3 k \over (2 \pi)^3 2 \omega} \, {4 \pi \alpha \over 3} \, \left ( \sum_{\rm pol} \left | \langle nP | \vec r | 1S \rangle \right |^2 \right ) \times \nonumber \\  &&{\Delta_n^2 \over \gamma^2}  \, 2\pi \, \delta \left [ \omega (1- \cos \theta) - {\Delta_n \over \gamma} \right ]~.
\label{gen} 
\eea
In this expression   $\omega$ and $\vec k$ are the energy and the momentum of the photon in the bath and $\theta$ is the angle between the momenta of the photon and the ion in the CMB frame as well as the photon thermal distribution function $n_{\rm CMB}(\omega) =  [\exp (\omega/T)-1]^{-1}$ at $T=T_{CMB}=2.725\,$K, while the excitation energy $\Delta_n = \varepsilon (nP) - \varepsilon (1S)$ and the dipole matrix element $\langle nP | \vec r | 1S \rangle$ refer to the rest frame of the ion, and the sum runs over the polarizations of the $nP$ state. Finally $\gamma = E/M$ is the $\gamma$ factor for the ion in the CMB frame. After angular integration the expression (\ref{gen}) takes the simple form:
\bea
&&\Gamma_{en}={2 \alpha \over 3} \, \left ( \sum_{\rm pol} \left | \langle nP | \vec r | 1S \rangle \right |^2 \right ) \, {\Delta_n^2 \over \gamma^2} \int_{\Delta_n \over 2 \gamma}^\infty n(\omega) \, d\omega \nonumber \\
&& = {8 \alpha \over 3} \, \left ( \sum_{\rm pol} \left | \langle nP | \vec r | 1S \rangle \right |^2 \right ) \, T^3  \, \chi \left({\Delta_n \over 2 \gamma T} \right ), 
\label{genf}
\eea
where 
\be
\chi (z) = -z^2 \, \ln \left ( 1- e^{-z} \right )~.
\label{fz}
\ee
It can be noticed that the lower limit in the $\omega$ integral is the lowest photon energy at which the excitation to the energy $\Delta_n$ is kinematically possible (in a head-on collision).

For a simple hydrogen-like one-electron ion the dipole matrix elements in Eq.(\ref{genf}) for discrete bound states can be readily found in terms of the rescaled Bohr radius $a_Z = a_B/Z = (m_e \alpha Z)^{-1}$, using the explicit form of the hydrogen wave functions:
\be
\sum_{\rm pol} \left | \langle nP | \vec r | 1S \rangle \right |^2={256 \, n^7 \over (n^2-1)^5} \, \left ( {n-1 \over n+1} \right )^{2n} \, a_Z^2~, ~~~~~~~(n \ge 2)
\label{dme}
\ee
with the excitation energy, clearly, expressed through the re-scaled Rydberg energy $\varepsilon_Z = Z^2 \, {\rm Ry} = m_e^2 \alpha^2 Z^2/2$ as $\Delta_n = (1-n^{-2}) \, \varepsilon_z$. 

Using Eq.(\ref{genf}), one can find the total excitation rate $\Gamma_{\rm e}= 1/\lambda_{\rm e }$ of the ion to bound states:  
\be
\Gamma_{\rm e} = {8 \alpha \over 3} a_Z^2 T^3 X \left ( {\gamma \over \gamma_Z} \right ) = {1 \over 2.46 \times 10^7 {\rm s}} \, X \left ( {\gamma \over \gamma_Z} \right ) \, \left (26 \over Z \right)^2~,
\label{gesn}
\ee
where 
\be
X(r)=256 \, \sum_{n=2}^\infty \, { n^7 \over (n^2-1)^5} \, \left ( {n-1 \over n+1} \right )^{2n} \, \chi \left [ (1- n^{-2})/ r \right ]~,
\label{fzs}
\ee
with $\gamma_Z = \varepsilon_Z/(2T)=1.96 \times 10^7 \, (Z/26)^2$ being the characteristic value of the gamma factor in the present discussion.  
Function $X(r)$, computed numerically is shown in Fig.~1. The series in Eq.~(\ref{fzs}) converges quickly, and even the first term provides an excellent approximation to the sum.

Excitation of the states higher than $2P$ can result in the atomic cascade emission of a gamma ray with energy boosted from at least $\Delta_2 = (3/4) \, \varepsilon_Z$. However the contribution of the higher excitations to the rate $\Gamma_{\rm e}$ is small. We, therefore, use the approximation that every excitation results in emission of a photon with energy  $(3/4) \, \varepsilon_Z = 6.90 \, (Z/26)^2\,$keV, which,  boosted to the laboratory frame, produces a uniform spectrum of gamma rays with energies up to $(3/4) \, \gamma \varepsilon_Z = 101 \, {\rm GeV} \, (Z/26)^4 \, (\gamma / \gamma_Z)$.

\begin{figure}
 \begin{center}
       \includegraphics[width=85mm]{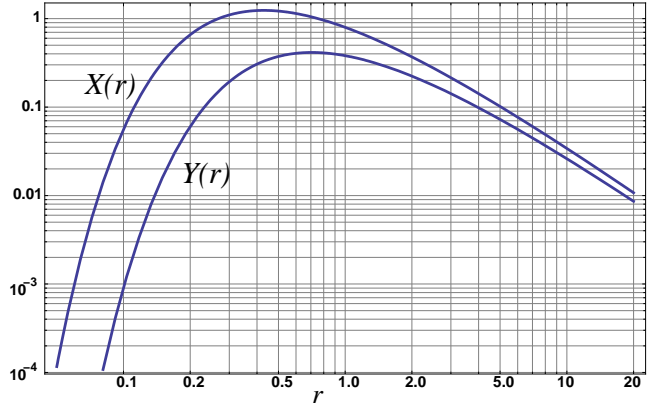}
\caption{
The functions $X(r)$ and $Y(r)$ describing the rates of excitation and ionization, respectively, versus the ratio $r=\gamma/\gamma_Z$. }
    \end{center}
  \label{fig:excitation}
\end{figure}

The rate of the ionization process in which the `active' ions are destroyed can be calculated similarly to the discrete excitation, in terms of the atomic ionization function
\be
S(\Delta)=\int {d^3 p \over (2 \pi)^3} \,  \delta \left( \Delta - {p^2 \over 2 m_e} - \varepsilon_Z \right ) \,   \left | \langle \vec p | \vec r | 1S \rangle \right |^2~,
\label{sf}
\ee
where $|\vec p \rangle$ is the state in continuum in the Coulomb potential whose momentum at infinity is asymptotically $\vec p$.  This function is well known in the theory 
of hydrogen atom, and can be read off from , {\em e.g.} Ref.~\cite{holt69}: $S(\Delta) = a_Z^2 \, \varepsilon_Z \, s(\Delta/\varepsilon_Z)$, where the dimensionless function $s(w)$ is given by 
\bea
&&s(w)= {128  \over w^3} \, {1 \over 1- \exp(-2 \pi/\sqrt{w-1})} \times \nonumber \\ 
&&\exp \left(-{2 \over \sqrt{w-1}} \, \arctan {2 \sqrt{w-1} \over 2-w} \right )~,
\label{s1s}
\eea
and where the branch of the arctangent must be chosen to take the values between $0$ and $\pi$.
Similar to Eq.(\ref{genf}), the ionization rate $\Gamma_{\rm i}$ in the CMB thermal bath, including the ionization to all energies above the threshold can be written as
\bea
\label{gis}
&&\Gamma_{\rm i} =\frac{1}{\lambda_{\rm i}}= {8 \alpha \over 3} \, T^3 \int \, S(\Delta) \, \chi \left( {\Delta \over 2 \gamma T} \right ) \, {d \Delta \over \Delta^2} =  \\ \nonumber
&&{8 \alpha \over 3} \, a_Z^2 \, T^3 \, Y\left ( {\gamma \over \gamma_Z} \right )={1 \over 2.46 \times 10^7 {\rm s}} \, Y \left ( {\gamma \over \gamma_Z} \right ) \, \left (26 \over Z \right)^2, 
\eea
where 
\be
Y(r)=\int_{w=1}^{\infty} \, s(w) \, \chi \left ( {w \over r} \right ) \, {dw \over w^2}~.
\label{gor}
\ee
The ionization function $Y(r)$ is readily calculated numerically, and the plot is shown in Fig.~1. Clearly, the ratio of the excitation and ionization factors determines the average number of gamma photons emitted by an ion during its life time $\tau_i=1/\Gamma_{\rm i}$, $\eta = X/Y$. The plot for this ratio is shown in Fig.~2.

\begin{figure}
 \begin{center}
       \includegraphics[width=85mm]{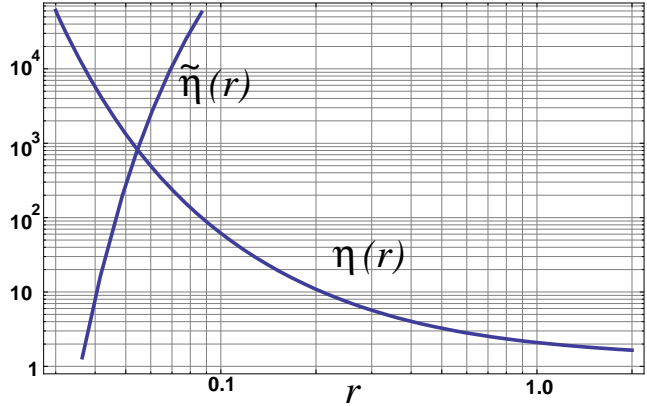}
\caption{
Two of the functions that determine the average number of gamma photons generated during the lifetime of an ion, $N_\gamma = \min \{ \eta,  \tilde{\eta},\hat{\eta} \}$. 
 }
    \end{center}
  \label{fig:ngamma}
\end{figure}

In addition to interactions with CMB, ionization can occur due to interactions with extragalactic background light (EBL), with energies between $ 10^{-3}$eV and 1~eV in the laboratory frame. Since all of these photons have relatively high energies, they ionize the ions and 
terminate the cycle of absorptions and emissions. The cross section for the relevant energy range is $\sigma \approx \sigma_{\rm T}=0.6$~barn, and the number density can be integrated numerically using Ref.~\cite{EBL}.  There are non-negligible uncertainties 
in the level of EBL, so that $\int dE (dn_{_{\rm EBL}}/dE)$ varies in the range $( 0.5 - 1.1) \, {\rm cm}^{-3}$, depending on a model~\cite{EBL}.  For a mid-range reference value of $0.8 \, {\rm cm}^{-3}$, 
the ionization by EBL photons occurs at a rate
$\Gamma_{\rm i;\, EBL} \approx 1/\lambda_{\rm i;\, EBL}\approx 1/({0.7 \, {\rm Mpc}}) $.
For a nucleus with a gamma factor $r\gamma_Z$ that passes distance $d$ (which can be the distance to the source, or the length of a diffusive path in turbulent magnetic field), the number of photons emitted in its lifetime cannot exceed $\eta = X/Y$, or $\tilde{\eta} = \lambda_{\rm i;\, EBL}/\lambda_{\rm e} $, or $\hat{\eta} = d/\lambda_{\rm e}$, so the total number of photons per nucleus is 
\be
N_\gamma = \min \{ \eta,  \tilde{\eta},\hat{\eta} \}.
\ee
As shown in Fig.~2, for a sufficiently distant source, the number of photons is maximized around $r=0.055$, which corresponds to the photon energy around $8$~GeV.

According to Eq.(\ref{gis}), the ion lifetime $\tau_i$ depends on its gamma factor. Using the calculated function $Y$ (Fig.~1), we estimate that this lifetime is minimal at $\gamma \approx 0.775 \gamma_Z$, where $Y=Y_{\rm max}  \approx 0.457$, which for $Z=26$ corresponds to $\tau_i \approx 5.4 \times 10^7\,$s $\approx 1.7\,$yr and $\gamma \approx 1.14 \times 10^7$, which gives the upper bound of the gamma photon spectrum at $E_\gamma \approx  79\,$GeV. 
The phenomenological significance of the discussed effect depends on the relation between the lifetime of the ion $\tau_i$ and the duration $\tau_N$ of the process in which heavy nuclei are generated and destroyed or decelerated. If $\tau_N \gg \tau_i$, the flux of the ions $F_i$ and the flux of the gamma photons $F_\gamma$ are related as
\be
F_\gamma = N_\gamma \, F_i 
\label{ff}
\ee

In the opposite case,  $\tau_N \ll \tau_i$, the rate of emission of the gamma photons exponentially decays and their flux is given by
$F_\gamma = \Gamma_{\rm e} \, e^{-\Gamma_{\rm i} t} \, \int F_i \, dt$, 
where the integral is over the duration of the burst flux of ions. However, for the mechanism in question, the time scale for creation  of the one-electron ions is determined by the beta decay times $\tau_\beta \sim 1 \div 10^4\,$s, dilated by the factor $\gamma$, which results in the effective emission times for the ions that are longer than $\tau_i$. Hence, in our subsequent estimates of the gamma-ray flux we will use Eq.(\ref{ff}).

We now compare our predictions with the data for diffuse background and for point sources.  Diffuse background is probably dominated by the 
nuclei diffusing in the magnetic field of our Milky Way galaxy.   The flux of nuclei at the relevant energy $E_N\sim 10^{17}$~eV can be 
inferred from the data~\cite{Bluemer:2009zf}: 
\be
F_{\rm Fe} = 2\times 10^{-14} {\rm cm}^{-2} {\rm s}^{-1} {\rm sr}^{-1}. \label{nuclearflux}
\ee 
This implies the local density of nuclei $n_{\rm Fe} = 1.3\times 10^{-23} {\rm cm}^{-3}$.  To translate this into the photon flux, one has to make assumptions about the 
largely unknown geometry of the diffusive magnetic fields in our galaxy.  The simplest estimate, assuming a spherical distribution of diffusing nuclei of the size $L$, in which 
each nucleus spends time $\tau$ yields diffuse photon flux at energies (5-10)~GeV
\be
F_\gamma = 10^{-11} \left( \frac{N_\gamma}{10^3} \right)
 \left(\frac{L}{100 \rm kpc} \right)  \left(\frac{10^6 \rm yr}{\tau} \right)
  {\rm cm}^{-2} {\rm s}^{-1} {\rm sr}^{-1}. 
\ee
This is well below the diffuse background reported by Fermi~\cite{Abdo:2010nz}.

We now discuss the predictions for point sources. One particularly interesting candidate is Centaurus A  (NGC~5128), which is the closest active galaxy at distance $d=3.4$~Mpc.  
A cluster of UHECR events in PAO data appears to favor Cen A as the origin of UHECR~\cite{Abreu:2011vm}. At lower energies, the galactic magnetic 
fields isotropize the arrival directions.   It is possible that all or most of UHECR come from a single nearby source~\cite{Cen_A}, or from galactic sources~\cite{Calvez:2010uh}.  
As an example, we apply our discussion to Cen A.  The observed flux in Eq.(\ref{nuclearflux}) can be used as an upper limit on the Cen A contribution.  If the 5--10~GeV shoulder that we have described was ruled out by the data, one could set a stronger upper limit on nuclear acceleration in Cen A.  
The emission of gamma rays is possible in a narrow band of energies, and the flux is enhanced around $r=0.055$, as shown in Fig.~2.  The dominant gamma-ray energy is $r \gamma_Z (7 \, \rm keV) = 8$~GeV.   In this case the expected flux of 8~GeV photons is 
$ F_{\rm Cen\, A} \sim 4\pi \, F_{\rm Fe} N_\gamma $, where we take $N_\gamma = 10^3$, consistent with Fig.~2.  
This flux is in agreement with the data, as shown in Fig.~3.  The spectral feature at 5--10~GeV may, in fact, result from the effect we have described.  However, there are good reasons to believe that 
local, Milky Way sources are responsible for most of the nuclear flux in this energy range~\cite{Bluemer:2009zf}.   

\begin{figure}
 \begin{center}
       \includegraphics[width=90mm]{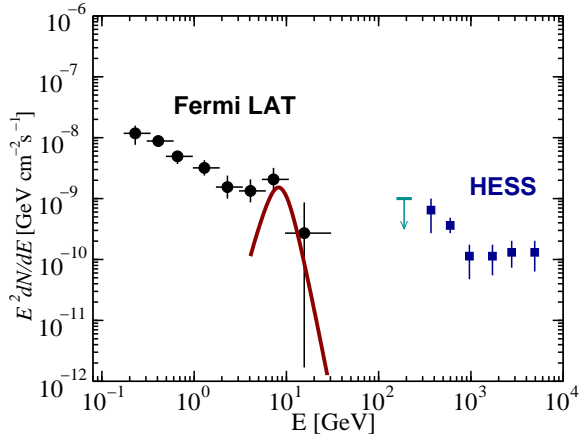}
\caption{
Signature of nuclear emission for Cen A (solid line) and the data from Fermi~\cite{Fermi} and HESS~\cite{Aharonian:2009xn,Aharonian:2005ar}. The flux normalization corresponds to 
the observed flux $F_{\rm Fe}(E  \gtrsim10^{17}{ \rm eV})=2\times 10^{-14} {\rm s}^{-1} {\rm cm}^{-2} {\rm sr}^{-1}$~\cite{Bluemer:2009zf}. Based on this comparison alone, the upper limit on 
Cen A luminosity in nuclei with energies above $10^{17}$eV is $4\times 10^{42}$erg/s, close to the observational upper limit. }
    \end{center}
  \label{fig:spectrum}
\end{figure}

Using the data from Fermi and other gamma-ray instruments, one may be able to identify powerful nuclear sources at large distances.  The shape of the ``shoulder'' due to cosmoluminescence may contain information about the range of isotopes among the nuclei.

In summary, we have described a new signature of astrophysical nuclear accelerators, which can be used to identify or constrain astrophysical sources of nuclei.

The authors thank  J.~Beacom, C.~Dermer, W. Essey, P.~M\'esz\'aros, and S.~Razzaque for discussions. A.K. and  M.B.V. were supported in part by DOE grants DE-FG03-91ER40662 and DE-FG02-94ER40823, respectively. The authors appreciate hospitality of the Aspen Center for Physics, supported by NSF grant 1066293, where this work was done.


\begin{thebibliography}{99}

\bibitem{Bluemer:2009zf}
  J.~Bluemer, R.~Engel, J.~R.~Hoerandel,
  Prog.\ Part.\ Nucl.\ Phys.\  {\bf 63} (2009) 293-338.

\bibitem{Abraham:2009dsa}
  J.~Abraham {\it et al.}  [The Pierre Auger Observatory Collaboration],
  arXiv:0906.2319; 
  Phys.\ Rev.\ Lett.\  {\bf 104} (2010) 091101;  
%
  A.~V.~Glushkov {\em et al.},
  JETP Lett.\  {\bf 87} (2008) 190;   
%
  R.~U.~Abbasi {\it et al.}  [HiRes Collaboration],
  Phys.\ Rev.\ Lett.\  {\bf 104} (2010) 161101.

\bibitem{signatures}
  L.~A.~Anchordoqui {\em et al.}, 
  Phys.\ Rev.\ Lett.\  {\bf 98} (2007) 121101; 
  L.~A.~Anchordoqui {\em et al.}, 
  Phys.\ Rev.\  D {\bf 75} (2007) 063001; 
  L.~A.~Anchordoqui {\em et al.}, 
  Astropart.\ Phys.\  {\bf 29} (2008) 1-13;
  K.~Murase, J.~F.~Beacom,
  Phys.\ Rev.\  {\bf D81}  (2010) 123001; 


\bibitem{survival}
  C.~D.~Dermer,
in Proceedings of 30th ICRC,  
  arXiv:0711.2804 [astro-ph]; 
  K.~Murase {\em et al.}, 
  Phys.\ Rev.\  D {\bf 78} (2008) 023005;
  X.~-Y.~Wang, S.~Razzaque, P.~M\'esz\'aros,
  Astrophys.\ J.\  {\bf 677} (2008) 432-440;
  A.~Pe'er, K.~Murase, P.~M\'esz\'aros,
  Phys.\ Rev.\  {\bf D80} (2009) 123018; 
  K.~Murase  {\em et al.}, 
  arXiv:1107.5576 [astro-ph.HE].

\bibitem{Essey:2009zg}
  W.~Essey and A.~Kusenko,
  Astropart.\ Phys.\  {\bf 33} (2010) 81;
  W.~Essey {\em et al.}, 
Phys. Rev. Lett. {\bf 104}  (2010) 141102; 
  W.~Essey {\em et al.}
  Astrophys.\ J.\  {\bf 731} (2010) 51.

\bibitem{Ando:2010rb}
  S.~Ando, A.~Kusenko,
  Astrophys.\ J.\  {\bf 722} (2010) L39; 
  W.~Essey, S.~Ando, A.~Kusenko,
  Astropart. Phys. {\bf 35} (2011) 135.


\bibitem{Bahcall:1961zz}
  J.~N.~Bahcall,
  Phys.\ Rev.\  {\bf 124} (1961) 495-499.
  

\bibitem{holt69}
A.~R.~Holt, J. Phys. B: At. Mol. Phys. {\bf 2} (1969) 1209.

\bibitem{EBL}
  F.~W.~Stecker, M.~A.~Malkan, S.~T.~Scully,
  Astrophys.\ J.\  {\bf 648} (2006) 774-783;
  A.~Franceschini, G.~Rodighiero, M.~Vaccari,
  Astron.\ Astrophys.\  {\bf 487} (2008) 837;
 R.~C.~Gilmore et al., MNRAS 399 (2009) 1694; 
  J.~D.~Finke, S.~Razzaque, C.~D.~Dermer,
  Astrophys.\ J.\  {\bf 712} (2010) 238; 
  M.~Orr, F.~Krennrich, E.~Dwek,
  Astrophys.\ J.\  {\bf 733} (2011) 77.




\bibitem{Abdo:2010nz}
  A.~A.~Abdo {\it et al.} [ The Fermi-LAT Collaboration ],
  Phys.\ Rev.\ Lett.\  {\bf 104} (2010) 101101.


\bibitem{Abreu:2011vm}
  P.~Abreu {\it et al.}  [The Pierre Auger Collaboration],
  JCAP {\bf 1106} (2011) 022.

\bibitem{Cen_A}
  C.~D.~Dermer {\em et al.}, 
  New J.\ Phys.\  {\bf 11} (2009) 065016; 
  M.~Kachelriess, S.~Ostapchenko, R.~Tomas,
  New J.\ Phys.\  {\bf 11} (2009) 065017; 
 L.~I.~Caramete {\em et al.}, 
  arXiv:1106.5109 [astro-ph.HE].
  A.~M.~Taylor, M.~Ahlers and F.~A.~Aharonian,
  arXiv:1107.2055 [astro-ph.HE].

\bibitem{Calvez:2010uh}
  A.~Calvez, A.~Kusenko, S.~Nagataki,
  Phys.\ Rev.\ Lett.\  {\bf 105} (2010) 091101.
\bibitem{Fermi}
  A.A. Abdo {\it et al.}  [Fermi Collaboration],
  Astrophys.\ J.\  {\bf 719} (2010) 1433;
Astrophys.\ J.\  {\bf 720} (2010) 912. 

\bibitem{Aharonian:2009xn}
  F.~Aharonian {\it et al.} [HESS Collaboration],
 Astrophys. J. Lett., {\bf 695} (2009) L40.

\bibitem{Aharonian:2005ar}
  F.~Aharonian {\it et al.}  [H.E.S.S. Collaboration],
  Astron.\ Astrophys.\  {\bf 441} (2005) 465.



\end{thebibliography}

\end{document}